**Mathematical Modeling of HDV RNA, HBV DNA, and HBsAg Dynamics during Lonafarnib-Based Therapy: Insights from the LOWR HDV-1 Study**

Adquate Mhlanga[1], Louis Shekhtman[1,2], Rami Zakh[3,4], Sarah Duehren[1], Ashish Goyal[1], Alexander Churkin[4], Vladimir Reinharz[5], Danny Barash[3], Jeffrey Glenn[6], Ohad Etzion[7], Scott J. Cotler[1], Cihan Yurdaydin[8], Harel Dahari[1*]

[1]The Program for Experimental and Theoretical Modeling, Division of Hepatology, Department of Medicine, Stritch School of Medicine, Loyola University Chicago, Maywood, Illinois, USA; [2]Department of Information Science, Bar-Ilan University, Israel; [3]Department of Computer Science, Ben-Gurion University, Beer-Sheva, Israel; [4]Department of Software Engineering, Sami Shamoon College of Engineering, Beer-Sheva, Israel; [5]Department of Computer Science, Universit´e du Qu´ebec ´a Montr´eal, Montr´eal, Canada; [6] Division of Gastroenterology and Hepatology, Departments of Medicine, Microbiology & Immunology, Stanford School of Medicine, Stanford, California, USA; [7]Department of Gastroenterology and Liver Diseases, Soroka University Medical Center, Beer-Sheva, Israel; [8]Department of Gastroenterology and Hepatology, Koç University Medical School, Istanbul, Turkey.

*Corresponding author: hdahari@luc.edu (H. Dahari)

**Financial support:** This work was supported in part by a grant from Binational Science Foundation grant number 207745, and US NIH grant R01AI146917. The funders had no role in study design, data collection and analysis, decision to publish, or preparation of the manuscript.

**Data availability statement:** Data is available on request from the authors.

,,

## Abstract

Lonafarnib (LNF) is an investigational drug targeting hepatitis delta virus (HDV) but not hepatitis B virus (HBV), providing a unique opportunity to model HDV kinetics and how changes in HDV affect HBV. We performed a detailed kinetic analysis and developed a mathematical model to explain serum HBV DNA, HDV RNA and hepatitis B surface antigen (HBsAg) kinetics in 15 HBV/HDV coinfected patients receiving LNF-based treatment. After a delay of 0-2 days, patients experienced a rapid 1st-phase HDV-decline followed by either a viral plateau, 2nd slower-decline phase, or viral breakthrough (VB). LNF monotherapy led to a flat-partial-response (often followed by VB), while LNF combination therapy with ritonavir or pegylated interferon-α (PEG-IFN$\alpha$) was associated with a biphasic HDV decline (without VB). All treatments except LNF+PEG-IFN$\alpha$ had at least one patient experiencing an increase in HBV on-treatment. Our model successfully reproduced the observed HDV and HBV kinetics. We estimated an HDV RNA half-life of 1.26 days [95% confidence interval, CI: 1.05–1.47] in serum and treatment efficacy of 94% in inhibiting HDV RNA production across all treatments [95% CI: 89%-97%], as reflected by the 1$^{st}$ phase HDV decline. The 2nd phase of HDV decline was explained by a time-dependent increase in efficacy, reaching a maximum of 98.9%. The model explained the increase in serum HBV DNA by a median 4-fold [interquartile range, IQR: 1 – 28] increase in HBV DNA production rate when HDV declined below an inhibitory threshold. The stability of serum HBsAg was explained by a constant number of HBsAg-producing cells.

# 1 Introduction

Previous modeling efforts, though limited, have provided insights into the interplay between hepatitis D virus (HDV) and hepatitis B virus (HBV). One of the earliest works in the field by de Sousa and Cunha (de Sousa and Cunha 2010) incorporated uninfected, HBV mono-infected, HDV mono-infected and HBV+HDV dually-infected hepatocytes alongside free virions to simulate coinfection and superinfection. They simulated viral load dynamics in response to therapies such as lamivudine and interferon; however, simulated responses did not match clinical observations (Rieche et al. 2018). Later on, Packer *et al*. (Packer et al. 2014) and Plaire *et al*. (Plaire, Jean Jules, and Bowong 2018) developed mathematical models specifically for coinfection with HBV and HDV, and performed steady state analysis. Packer *et al*. (Packer et al. 2014) suggested a 'theoretical' inhibitory effect of HDV on HBV via infected cell proliferation. Although useful, these models are not sufficiently robust to analyze and simulate specific anti-HDV treatments.

Similarly, Goyal and Murray developed complex multiscale models incorporating intracellular components (e.g., cccDNA, HDV RNA) that could aid investigation of not only anti-HDV treatment but also anti-HBV treatment and broad therapies such as interferon for which the mechanism of action is unclear (Goyal and Murray 2016). They simulated several kinetic patterns of serum HBV DNA and HDV RNA that are possible under anti-HDV treatment such as prenylation inhibitors (Goyal and Murray 2016). The complex nature of this model requires rich datasets which are generally available only in preclinical settings because it is not practical to obtain intracellular dynamics in patients. Therefore, we simplified the model of Goyal and Murray to study the clinical data from the Phase 2 pilot study 'Lonafarnib with and without Ritonavir in HDV – 1', also known as LOWR HDV-1 study (Yurdaydin et al. 2018). This clinical study evaluated the efficacy of lonafarnib (LNF), a prenylation inhibitor that disrupts HDV assembly, but does not affect the HBV lifecycle directly (Verrier et al. 2022; Glenn et al. 1992).

Our group has made previous efforts to model the LOWR HDV-1 study results (Zakh et al. 2021; Mhlanga et al. 2022). Zakh *et al*. (Zakh et al. 2021) developed a preliminary mathematical model that was later modified by Mhlanga *et al*.(Mhlanga et al. 2022), and mainly focused on HDV without including HBsAg kinetics. Here, we present an updated mathematical model to simultaneously characterize the kinetics of serum HDV RNA, HBV DNA, and HBsAg. The LOWR HDV-1 study offers a unique opportunity, as LNF specifically targets the HDV life cycle without directly inhibiting HBV replication. Modeling LNF monotherapy allows for assessment of how suppression of HDV RNA influences HBV DNA and HBsAg levels in patients coinfected with HBV and HDV.

In this study, we conducted a comprehensive kinetic analysis of serum HDV RNA, HBV DNA, ALT, HBsAg, and LNF concentration across all treatment arms, and developed a mathematical model that captures HBV and HDV viral kinetics. Collectively, these results enhance our understanding of HBV–HDV interplay under anti-HDV therapy and provide a quantitative framework to inform the design of more effective combination strategies.

# 2 Methods

## 2.1 Description of LOWR (Lonafarnib with or without Ritonavir) HDV-1 patient groups

The LOWR HDV-1 study (Yurdaydin et al. 2018), conducted at the University of Ankara Medical School in Turkey, was a single-center, phase 2 pilot trial that investigated the efficacy of LNF at various doses and in combination with ritonavir (RTV) and pegylated interferon (PEG-IFN$\alpha$) over 8-12 weeks. The study was approved by the Ankara University Ethics Committee and adhered to the ethical guidelines of the 1975 Declaration of Helsinki. The study design included seven groups of three patients each, namely, Group 1: LNF 200mg twice daily (BID) for 12 weeks; Group 2: LNF 300mg BID for 12 weeks; Group 3: LNF 100mg thrice daily (TID) for 8 weeks; Group 4: LNF 100mg BID + RTV 100mg once daily (QD) for 8 weeks; Group 5: LNF 100mg BID + PEG-IFN$\alpha$ 180$\mu$g once weekly (QW) for 8 weeks; Group 6: LNF 200mg BID + PEG-IFN$\alpha$ 180$\mu$g QW for 8 weeks; Group 7: LNF 300mg BID + PEG-IFN$\alpha$ 180$\mu$g QW for 8 weeks.

A total of 20 patients (14 males and 6 females) were enrolled with one patient from Group 3 re-enrolled in Group 7 after a six-month washout period (see **Fig. 1** in (Yurdaydin et al. 2018)). Five patients in Groups 6 and 7 discontinued treatment within four weeks due to poor tolerance. For this reason, viral load and pharmacokinetic data for these two groups are not available and they are excluded from our modelling efforts.

## 2.2 Patients' characteristics

Patients aged 18-65 years with documented HBsAg, anti-delta positivity for at least 6 months, and compensated liver disease were eligible after exclusion of other causes of chronic liver disease, **Table S1** (Yurdaydin et al. 2018). The study's precise inclusion and exclusion criteria were documented in **Fig. 1** and **Table S1** (Yurdaydin et al. 2018).

## 2.3 Measurement of viral load

HDV RNA was measured as previously described in (Koh et al. 2015; Yurdaydin et al. 2018) with a lower limit of quantification (LLOQ) of 70 IU/mL and a lower limit of detection (LoD) of 50 IU/mL (Yurdaydin et al. 2018). HBV DNA was measured using the CobasTaqMan HBV test (Roche Molecular Systems, Inc), which can detect HBV DNA in the range of 20-1.7×$10^8$ IU/mL. HBsAg was quantified using the Architect HBsAg assay with LLOQ of 0.05 IU/ml (Abbott Diagnostics, Germany).

## 2.4 Drug quantification

Liquid chromatography/tandem mass spectrometry (LC-MS/MS) methods were utilized to determine serum concentrations of LNF and RTV. The analysis range for LNF and RTV was 1-2500 ng/mL.

## 2.5 Statistical analysis

We analyzed the association between LNF dose, viral kinetics, and estimated model parameters using non-parametric tests. Fisher's exact test was used to compare parameters between biphasic and flat partial responders, while Kruskal-Wallis Test was used to test the association between estimated parameter values and doses. Mann-Whitney U test was employed to assess the association between estimated parameter values and treatment type (combination therapy vs. monotherapy). Alanine aminotransferase (ALT) and LNF plateau levels were analyzed using linear regression from day 2 until the end of treatment (EOT) or beginning of HDV breakthrough. Linear regression was employed to assess whether the observed kinetics were consistent with a plateau by testing if the estimated slope was statistically different from zero. In all analyses, we considered a p-value of $p<0.05$ as statistically significant (IBM SPSS version 28.0.1.1). R version 4.3.2 along with lixoftConnectors (v2023.1) and Rsmlx (v2023.1.5), were used to interface with Monolix and compute confidence intervals for population parameter estimates using bootstrap approach with 100 replicates.

## 2.6 Model description

Inspired by our previous work (Mhlanga et al. 2022), we present a new mathematical model that incorporates HBsAg dynamics (**Fig. 1**). The new model consists of 3 ordinary differential equations (ODEs) given by

$$\frac{dD}{dt} = (1-\varepsilon)p_D I_0 e^{-g(t-\tau)} - cD$$

$$\frac{dB}{dt} = p_B I_0 \left(1 + \left(\frac{\kappa}{D+LoD}\right)^n\right) - cB \qquad \text{(Eq. 1)}$$

$$\frac{dH}{dt} = p_H I_0 - c_H H$$

where *D*, *B* and *H* represent serum HDV RNA, serum HBV DNA and serum HBsAg, respectively. For simplicity, we assumed that cells co-infected with HBV and HDV ($I_0$) stay at a constant level during therapy. The per infected cell production rates of HDV RNA and HBsAg were denoted by $p_D$ and $p_H$, respectively.

Since HDV has been shown to induce inhibitory effects on the HBV lifecycle (Lucifora et al. 2023; Lütgehetmann et al. 2012), we assumed that the per infected cell production rate of HBV DNA is HDV RNA dependent and is given by $p_B\left(1+\left(\frac{\kappa}{D+LoD}\right)^n\right)$. Here, LOD denotes the lower limit of detection for HDV RNA and was incorporated to reflect the practical constraints of the measurable range used in fitting the data. The parameter *n*, where $n > 1$, governs the rate of lifting of the inhibitory pressure of HDV on HBV and the parameter $\kappa$ represents the serum HDV RNA threshold that triggers an increase in HBV DNA production rate. We hypothesized that both parameters *n* and $\kappa$ could be patient-specific, governed by host and immune factors. In the absence of HDV RNA (i.e., HBV mono-infected individuals), the production rate of HBV DNA collapses to $p_B\left(1+\left(\frac{\kappa}{LoD}\right)^n\right)$. We also assumed that HDV RNA levels at baseline ($D_0$) were much higher than $\kappa$. Under this assumption, the production rate of HBV DNA at baseline can be approximated as $p_B$. We further assumed that the clearance rate, $c$, is the same for HDV and HBV since both share the same envelope proteins. Additionally, $c_H$ represented the clearance rate of serum HBsAg.

Serum HDV RNA, HBV DNA and HBsAg were found to remain stable over extended periods of time, typically weeks to months, with only minor fluctuations in the absence of treatment (Mederacke et al. 2010). The stable levels of serum HDV RNA, HBV DNA and HBsAg implies that viral production and clearance rates are in balance before treatment leading to the steady state conditions described below.

$$p_D = \frac{c\ D_0}{I_0} \qquad \text{(Eq. 2)},$$

$$p_B = \frac{c\ B_0}{I_0\left(1+\left(\frac{\kappa}{D_0+LoD}\right)^n\right)} \qquad \text{(Eq. 3)},$$

$$p_H = \frac{c_H\ H_0}{I_0} \qquad \text{(Eq. 4)}$$

where $B_0$ and $H_0$ are baseline HBV DNA and HBsAg levels, respectively, at the onset of treatment.

We further assumed that treatment blocks HDV RNA production with an efficacy of $\varepsilon$ (0 ≤$\varepsilon$ ≤ 1). Parameter $g$ is the additional time-dependent treatment inhibitory effect in blocking HDV RNA production, reflecting recent in vitro findings that LNF progressively increases the concentration of edited, non-infectious HDV genomes over time (Verrier et al. 2022). This term is reminiscent of previous modeling efforts where antiviral efficacy was assumed to increase over time (Kadelka, Dahari, and Ciupe 2021; Reinharz et al. 2021; Dahari, Sainz, et al. 2009).

We assumed that there is a delay ($\tau$) before serum HDV RNA declines from its baseline levels after treatment initiation, representing the pharmacological delay between treatment administration and the onset of measurable antiviral activity. During the delay period ($t \leq \tau$), there is no treatment effect (i.e. $g$ = $\varepsilon$ = 0), and viral dynamics remain at their pre-treatment steady state (**Eq. 2-4**). Once $t > \tau$, treatment effects begin and the system follows **Eq. 1**, where the term ($t - \tau$) ensures that the time-dependent antiviral effect accumulates only after the onset of drug action.

A recent in vitro study also demonstrated the existence of non-infectious HDV particles (Verrier et al. 2022). Motivated by these findings, we also considered an extension to our model distinguishing between infectious and non-infectious HDV RNA particles (see Supplementary Material C).

### 2.7 Parameter estimation and data fitting

The constant number of infected cells ($I_0$) was set to $1 \times 10^6$ cells/mL (Shekhtman et al. 2024). LoD was set to 1 IU/mL and can be adjusted to match each assays specific detection limit in future studies. The clearance rate of serum HBsAg was fixed at $c_H$ = 0.53 days$^{-1}$ as previously determined (Shekhtman et al. 2020).

Due to identifiability issues with the pharmacological delay ($\tau$) , we chose to fix it for each patient (**Table S3**). We manually estimated $\tau$ using the kinetic data and assumed that HDV RNA fluctuations within 0.2 $\log_{10}$ IU/mL from baseline reflect stable HDV RNA levels.

The model was fit for each patient up to HDV breakthrough, defined as an increase ≥ 0.5 $\log_{10}$ IU/mL, or the maximal treatment duration in the absence of HDV breakthrough leading to different lengths of fit for each patient.

The value (1-$E_{max}$) denotes the maximum effectiveness in blocking viral production on-treatment, and since we assumed that the efficacy increases with time, it was evaluated at the last observed time point used in the fitting (i.e., $t_{max}$). It is defined by the function:

$$E_{max} = (1 - \varepsilon)e^{-g(t_{max}-\tau)} \quad \text{(Eq. 5)}$$

The corresponding maximum efficacy of inhibition of HDV RNA production is given by 1-$E_{max}$.

We employed a population level approach to estimate parameters. In this approach, a nonlinear mixed-effects (NLME) model is used where model parameters are represented as $\theta_i = \mu e^{\eta_i}$, and $\mu$ represents the fixed effects describing the median of the population while $\eta_i$ denotes the random effects account for inter-individual variability (IIV). Model parameters were estimated using a maximum-likelihood method implemented in MONOLIX version 2023R1 [http://software.monolix.org](Lixoft 2024). A detailed explanation is available in **Supplementary material B**. In general, lower relative standard error (RSE) values indicate better precision, while higher values indicate greater uncertainty (Lixoft 2024).

The half-life of HDV and HBV was derived from the population estimate parameter c as $t_{1/2} = \ln(2)/c$, and the corresponding confidence interval was obtained using the delta method (Lavielle 2014; Bonate 2011).

### 2.8 Viral Kinetic Analysis

All patients were classified into viral kinetic patterns according to the number of identified phases (or slopes). A phase change was defined as a 2-fold difference in the slope between time points. A monophasic (or single-phase) pattern was defined by a single phase of decline. A biphasic (or two-phase) pattern was defined by a first phase of decline at least two-fold larger than the second phase decline rate. A flat partial response was defined by a first phase of viral decline following a plateau (zero-slope) at a lower viral load from baseline.

# 3 Results

## 3.1 Kinetic Analyses

The kinetics of HDV RNA, HBV DNA, HBsAg, ALT for 15 patients were placed into five groups of three patients each based on the treatment and dosage they received (**Fig. S1**). We present the kinetics for all patients until either the point of HDV RNA breakthrough (a confirmed rebound in HDV RNA after an initial decline while on therapy) or the end of the treatment (**Table 1**). ALT and LNF kinetics are provided in **Supplementary Material A**

### 3.1.1 HDV Kinetics

The median baseline serum HDV RNA level was 6.08 $\log_{10}$ IU/mL (IQR: 1.51) (**Table 1**). In ~67% of patients (n=10), HDV RNA remained unchanged for 1-2 days after the initiation of therapy, indicating a delay before response (**Fig. S1**). The remaining 5 patients (~33%) had an immediate HDV RNA decline.

Three patterns of HDV RNA decline were observed (**Table 1**, **Fig. S1**): (i) a flat partial response (n=8), including 4 patients who later experienced a breakthrough (**Fig. 2**, **Table S2**); (ii) a biphasic decline with two distinct rates (n=6), without any breakthroughs; and (iii) a monophasic decline (n=1), followed by a viral breakthrough. All patients had a similar rapid first phase decline with a median rate of 0.14 $\log_{10}$/day (IQR: 0.06) lasting for a median of 7 days (IQR:

7). Patients with a biphasic pattern then experienced a second, slower phase of decline with a median rate of 0.04 $\log_{10}$/day (IQR: 0.02) (**Fig. S1, Table 1**). In contrast, the viral load in flat partial response patients plateaued after the initial rapid decline until the end of treatment or viral breakthrough (**Fig. 2, Table 1**).

Next, we investigated if the type of therapy (combination therapy or monotherapy) was associated with the kinetics of HDV RNA decline. Six patients received LNF in combination with either RTV or PEG-IFN$\alpha$, and 5 of these 6 patients had a biphasic viral kinetic pattern. In contrast, 7/9 patients receiving LNF monotherapy had a flat partial response (**Table 1**). We found a significant association between combination therapy and a biphasic pattern (p=0.023). There was no association between LNF plateau concentrations levels and HDV RNA kinetic patterns (p=0.292).

### 3.1.2 <u>HBV Kinetics</u>

The median pre-treatment HBV DNA was 2.40 $\log_{10}$ IU/mL (IQR: 1.5) (**Table 1**). Three distinct serum HBV DNA kinetic patterns were observed; continuous undetectability (<20 IU/ml), detectable with no change (flat), and an increase. Three of 15 patients experienced no change in HBV DNA from baseline. Nine patients experienced an increase in HBV DNA during LNF treatment, with a median of 1.15 $\log_{10}$ IU/mL (IQR: 2.75) from baseline. Two patients remained HBV DNA undetectable throughout treatment, and one patient fluctuated between undetectability and 2 $\log_{10}$ IU/mL throughout treatment, ultimately returning to the baseline value at EOT (**Fig. S1**).

Two of 3 patients receiving LNF 200 mg BID had a flat response while the third patient showed an increase in HBV DNA starting on day 7. All patients receiving either LNF 300 mg BID or LNF + RTV exhibited an increase in HBV DNA. Patients receiving LNF 100 mg TID had mixed outcomes, with an increase in HBV DNA observed in 2 patients and 1 patient with continuous suppression until dropping out of the study on day 28. LNF + PEG-IFN$\alpha$ resulted in continuous suppression in 2 out of 3 patients while 1 patient showed a flat response. Overall, 50% of combination therapy cases and 66% of monotherapy cases had an increase in HBV DNA (**Table 1**,

**Figure 2**). The association between HBV DNA kinetics and LNF plateau concentrations levels did not reach statistical significance (p=0.06).

### 3.1.3 HBsAg Kinetics

HBsAg was measured in all patients, and they did not change throughout treatment (**Fig. S1**). The median baseline value of HBsAg was 3.84 $log_{10}$ IU/mL (IQR 0.44) (**Table 1**).

## 3.2 Modeling results and parameter estimates

The model (**Fig. 1** and **Eq. 1**) reproduced serum HDV RNA, HBV DNA and HBsAg kinetics in all patients, as shown in **Fig. 2**.

Using the population level fitting approach, we estimated parameters and corresponding inter-individual variability (IIV), as shown in **Table 2.** Estimated Parameters at the individual patient level are presented in **Table 3**. Model fits estimated baseline serum HDV RNA levels at 6.35 $log_{10}$ IU/mL [95% CI: 5.83 – 6.82]. After a delay of 0-2 days, HDV RNA began to decline, with the 1$^{st}$ phase of HDV RNA decline governed by the parameters *c* and $\varepsilon$. The clearance rate constant for both serum HDV RNA and HBV DNA, *c*, was estimated at 0.55 day$^{-1}$ [95% CI: 0.50 – 0.68], corresponding to HDV RNA and HBV DNA ($t_{1/2}$) of 1.26 days [95% CI: 1.05 – 1.47]. The efficacy, ε, in blocking HDV RNA production was found to be 94% [95% CI: 89% – 97%]. The second phase of HDV RNA decline in patients exhibiting a biphasic decline was attributed to the additional time-dependent inhibitory effect, *g*, on HDV RNA production, with an estimated value of 0.043 day$^{-1}$ [95% CI: 0.026 – 0.069]. For patients receiving monotherapy, parameter *g* tended to approach 0, indicative of a flat-partial response in HDV RNA, while combination therapy yielded higher *g* values, associated with a biphasic response (**Table 1**). Using **Eq. 5**, we were able to compute the maximal effectiveness, including the time dependent increase in inhibition of HDV RNA production in patients showing a flat partial response and a biphasic response, which were 98.84% (range: 93 – 99.78%) and 99.96% (range: 97.03 – 99.99%) respectively.

Our analysis yielded a baseline estimate of 2.33 $log_{10}$ IU/mL [95% CI: 1.77 – 2.96] for serum HBV DNA. A majority of patients (9/15) had an increase in HBV DNA during treatment which was explained by parameters *n* (steepness) and $\kappa$ (serum HDV RNA threshold triggering an increase in HBV DNA production rate). The two parameters $n$ and $\kappa$, were estimated to be 1.04 [95% CI: 1.00 – 2.91] and 2.78 $\log_{10}$ IU/mL [95% CI: 0.99 – 4.77] (**Table 3**), corresponding of a ~4-fold increase in the HBV DNA production rate when $\kappa$ was reached. In patients receiving LNF + PEG-IFN$\alpha$, serum HBV DNA remained unchanged which we attributed to an effect of PEG-IFNα treatment, which could suppress both HDV RNA and HBV DNA production.

HBsAg levels remained stable during the treatment period, with a median baseline of 3.75 $\log_{10}$ IU/mL [95% CI: 3.54 – 3.93].

We tested if parameters differed across treatment groups 1-5 and found no statistically significant differences for parameters $D_0$, $c$, $\epsilon$, $B_0$, $k$, $n$ and $H_0$ ($p > 0.05$, **Table S3**). Only parameter $\kappa$ exhibited significant variation between treatment groups ($p = 0.042$, **Table S3**). Furthermore, when comparing monotherapy versus combination therapy, only parameter g was significantly different between the 2 groups (i.e. g was higher in the combination therapy group, $p = 0.036$, **Table 3**).

We also examined an extended model (**Eq. S4**) that distinguishes between infectious and non-infectious HDV particles. Varying the assumed ratio (infectious vs non-infectious; 90% vs 10%, 50% vs 50%, and 10% vs 90%) had little effect on parameter estimates and did not change the results from the primary model (**Eq. 1**), see **Supplementary Material C** for full details.

# 4 Discussion

Here we developed a new model of LNF therapy for HDV and fit this model to kinetics of serum HDV RNA, HBV DNA and HBsAg in patients receiving either LNF monotherapy or LNF combination therapy with either RTV or PEG-IFNα from the LOWR-1 study (Yurdaydin et al. 2018). Distinct kinetic patterns were observed including declines in serum HDV RNA, increases in serum HBV DNA, and stable HBsAg levels in the same patient. Our model successfully reproduced the kinetic patterns of all three serum biomarkers including a biphasic decline of serum HDV RNA in a majority of patients receiving combination therapy (LNF with RTV or PEG-IFNα) and a flat partial response in a majority of patients receiving LNF monotherapy. The mathematical model provides a quantitative basis for understanding the antiviral response to LNF and why different treatment regimens produce distinct viral kinetic patterns.

The first phase efficacy of LNF in blocking HDV RNA production rate was estimated at 94% [95% CI: 89%–97%] and was similar between LNF monotherapy and combination therapy. It is worth noting that the addition of RTV or PEG-IFN$\alpha$ to LNF does result in a continued rise in treatment efficacy beyond the first phase, with maximum efficacy achieved by end of treatment. Interestingly, a considerably shorter delay before drug efficacy is apparent was observed for LNF (0-2 days), consistent with our previous estimate of 0.75 days [standard error 0.24](Koh et al. 2015). This delay is shorter compared to previous estimates for other HDV treatments such as PEG-IFN$\alpha$ (8.5 days) and nucleic acid polymers or NAPS (25 days) (Guedj et al. 2014; Shekhtman et al. 2020). The rapid effect of LNF could be because prenylation inhibitors act directly on the assembly and release of HDV, the stage closest to renewing virus in serum, hence their impact may be observed more rapidly compared to agents like PEG-IFN$\alpha$ or NAPs which operate at earlier stages in the HDV lifecycle or through indirect and complex processes. The clearance rate constant (*c*) was estimated at 0.55 [95% CI: 0.50 – 0.68] day$^{-1}$, corresponding to a half-life of serum HDV RNA of approximately 1.26 days [CI: 1.05 – 1.47], consistent with previous estimates of 1.65 days (Shekhtman et al. 2020).

A vital distinction between LNF monotherapy and combination therapy (LNF with either RTV or PEG-IFN$\alpha$) was captured by the parameter g, which represented the additional time-dependent inhibitory effect on HDV RNA production rate. In patients who received LNF monotherapy, g tended to approach 0 (**Table 1**, **Table S3**), consistent with a flat partial response in which HDV RNA declined initially but then plateaued. In contrast, patients who received combination therapy exhibited higher *g* values, corresponding to a sustained second phase of HDV RNA decline and a biphasic pattern. The addition of RTV or PEG-IFN$\alpha$ to LNF resulted in a continued rise in treatment efficacy beyond the first phase, with maximum efficacy achieved by end of treatment. We hypothesize that the biphasic pattern might be due to higher LNF concentrations during the second phase decline in patients receiving LNF+RTV compared to patients receiving LNF monotherapy (**Table S4**) or the antiviral effects of PEG-IFN$\alpha$ in patients receiving LNF+ PEG-IFN$\alpha$ (Yurdaydin et al. 2018; Giersch et al. 2023).

During anti-HDV treatment with LNF, an increase in serum HBV DNA was observed as serum HDV RNA was suppressed. Our model predicted that the decrease in serum HDV RNA resulting from inhibition of HDV RNA production was associated with a median 4-fold (range: 1-5347) increase in the HBV DNA production rate (**Table S5**). We hypothesize that the observed increase in HBV DNA on anti-HDV treatment reflects reduction in the inhibitory effects of HDV on the HBV lifecycle. This is consistent with the experimental findings that co-infection with HDV suppress chronic HBV in humans (Sureau and Negro 2016; Sausen et al. 2022) and HBV replication in humanized mice (Giersch et al. 2015; Tsuge et al. 2019; Lütgehetmann et al. 2012). Further experimental studies are warranted to confirm these findings and to investigate the mechanism.

The patient specific threshold for reversal of HDV mediated suppression of HBV replication, represented by parameter $\kappa$ , was estimated at a median HDV level of 2.72 [0.78 – 4.98] $\log_{10}$ IU/mL. Patients receiving LNF + PEG-IFNα exhibited lower $\kappa$ estimates compared to other treatment groups, suggesting that the immunomodulatory effects of PEG-IFNα on both HDV and HBV contribute meaningfully to differences in this parameter across treatment groups. Whether a stronger immune response controls the virus sufficiently to allow for a lower value of $\kappa$, while a weaker immune response results in higher values of $\kappa$, warrants further investigation.

Recent in vitro evidence suggests that LNF may affect virus infectivity in addition to blocking HDV production (Verrier et al. 2022), raising the question of whether such effects could influence our model conclusions. Therefore, we explicitly considered whether LNF could increase the proportion of non-infectious viral particles in addition to blocking HDV RNA production (see **Eq. S4**, **Fig. S3**). Since the proportion of non-infectious viral particles remains unknown, we incorporated different ratios of infectious and non-infectious virus (infectious vs non-infectious; 90% vs 10%, 50% vs 50%, and 10% vs 90%) though we did not consider changes in total HDV as has been done for hepatitis C virus (Goyal et al. 2017). We found minimal to no effects on the alternative model on our model conclusions, see **Supplementary material C.**

Our study has several limitations. First, we assumed that the half-life of serum HBV DNA is the same as the half-life of serum HDV RNA. A previous study provided evidence that the HBV DNA $t_{1/2}$ might differ between HBeAg-positive and HBeAg-negative HBV-mono-infected patients (Dahari, Cotler, et al. 2009). Since the majority of patients (n=12) in the LOWR-1 cohort were HBeAg-negative and had low baseline HBV DNA levels and no apparent decline during treatment, it was not possible to estimate the half-life of serum HBV DNA. Second, our model assumed a constant population of HBsAg producing hepatocytes. This assumption was motivated by the relatively stable HBsAg levels observed during therapy, which we interpreted as being consistent with minimal changes in the infected hepatocyte population, though it is also possible that viral spread is compensated for by cell death, resulting in a roughly constant infected cell population. Studies have shown that a limited number of cells are positive for HBsAg (Battistaella et al. 2026, Hercun et al. 2023). To remain conservative, we fixed this infected cell population at 10^6 cells/mL, which is reflective of the lower estimate of less than 10% HBsAg producing hepatocytes as recently done (Shekhtman et al. 2024). Third, while our model was able to explain HDV RNA and HBV DNA kinetics, it did not capture HDV breakthroughs (Table S2). Fourth, the proposed model is semi-mechanistic in nature which limits its universal application. Specifically, some parameters such as *g* capture the time-dependent additional blocking of HDV RNA production without modeling the underlying intracellular mechanisms driving this effect. Whether the parameter *g* reflects the loss of infected cells needs to be explored in the future when more data become available. Finally, we did not incorporate ALT or LNF concentrations into our model, though these did not appear to be correlated with any of the parameters we estimated (**Supplementary material A**).

Overall, our model captures the within-host dynamics of serum HDV RNA, HBV DNA, and HBsAg under LNF-based therapy. The model provides a detailed explanation for diverse viral kinetic patterns observed across the full LOWR-1 cohort. This framework serves as a good starting point to improve our understanding of HDV RNA suppression and an associated increase in HBV DNA in patients receiving an anti-HDV treatment which does not directly affect HBV . Further studies are needed to delineate the interplay between HDV and HBV at the molecular level, leveraging in silico, in vitro, and in vivo models.

## Figures

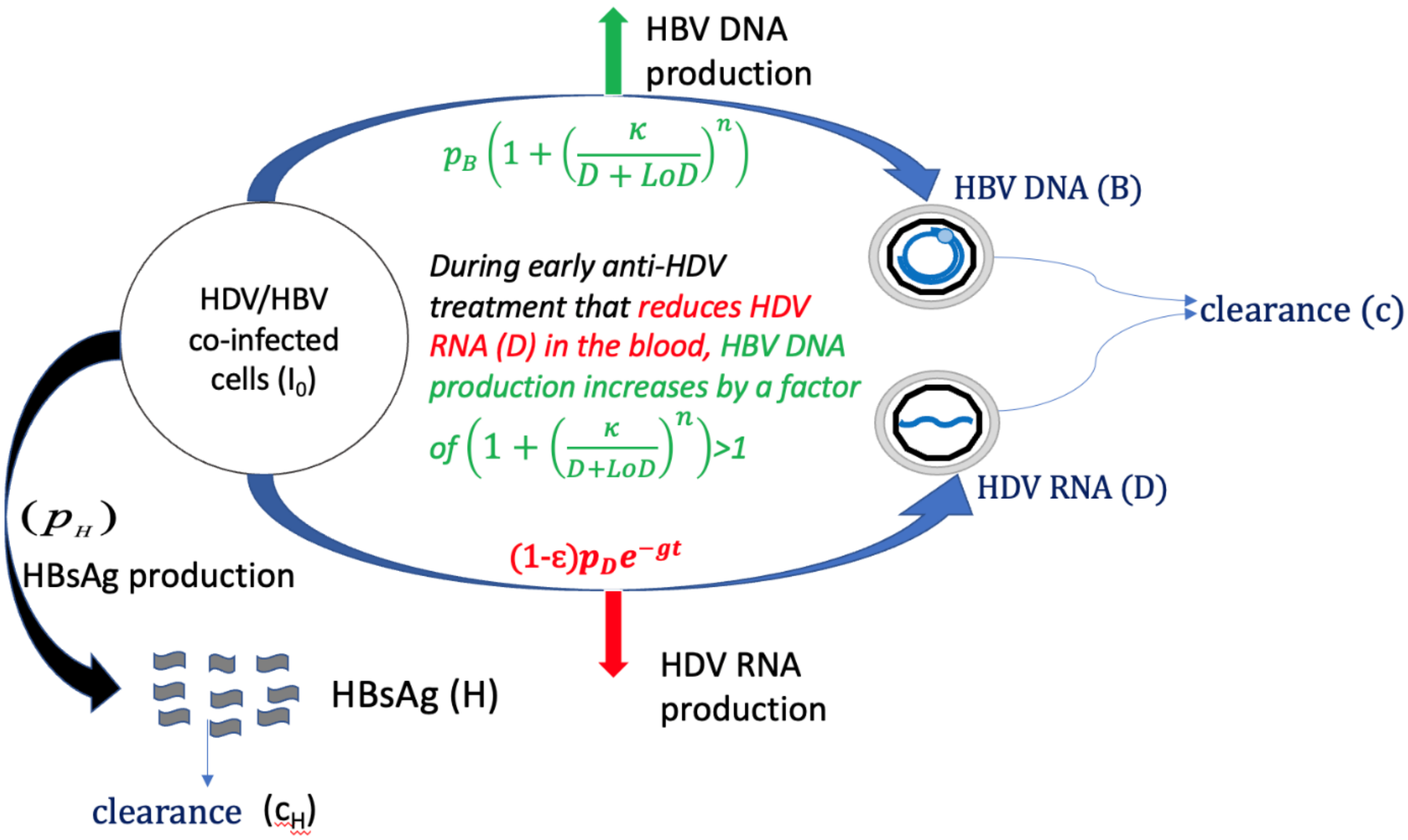

**Fig. 1**: The schematic diagram for our proposed model (**Eq. 1**). It shows the viral dynamics of HDV RNA (D), HBV DNA (B) and HBsAg (H) in serum. Parameters $D_0$, $B_0$, and $H_0$ represent serum HDV RNA, HBV DNA and HBsAg levels at the onset of treatment (not shown in Fig. 1). Parameters $p_D$, $p_B$, and $p_H$ denote the steady-state production rates of serum HDV RNA, HBV DNA, and HBsAg respectively. Parameter $\varepsilon$ represents the treatment efficacy in blocking HDV RNA viral production. Both serum HDV RNA and HBV DNA share the same clearance rate constant, c, while $c_H$ is the clearance rate of serum HBsAg. An additional treatment inhibitory effect in blocking HDV RNA production is represented by $g$. Parameter $n$ denotes the exponent in the process that governs the increase in the rate of HBV DNA production. Parameter $\kappa$ represents the threshold level for HDV RNA that triggers HBV DNA production increase, assuming $D_0 > \kappa$. Moreover, $I_0$ denotes cells that are co-infected by both HBV and HDV (i.e., HDV/HBV co-infected cells).

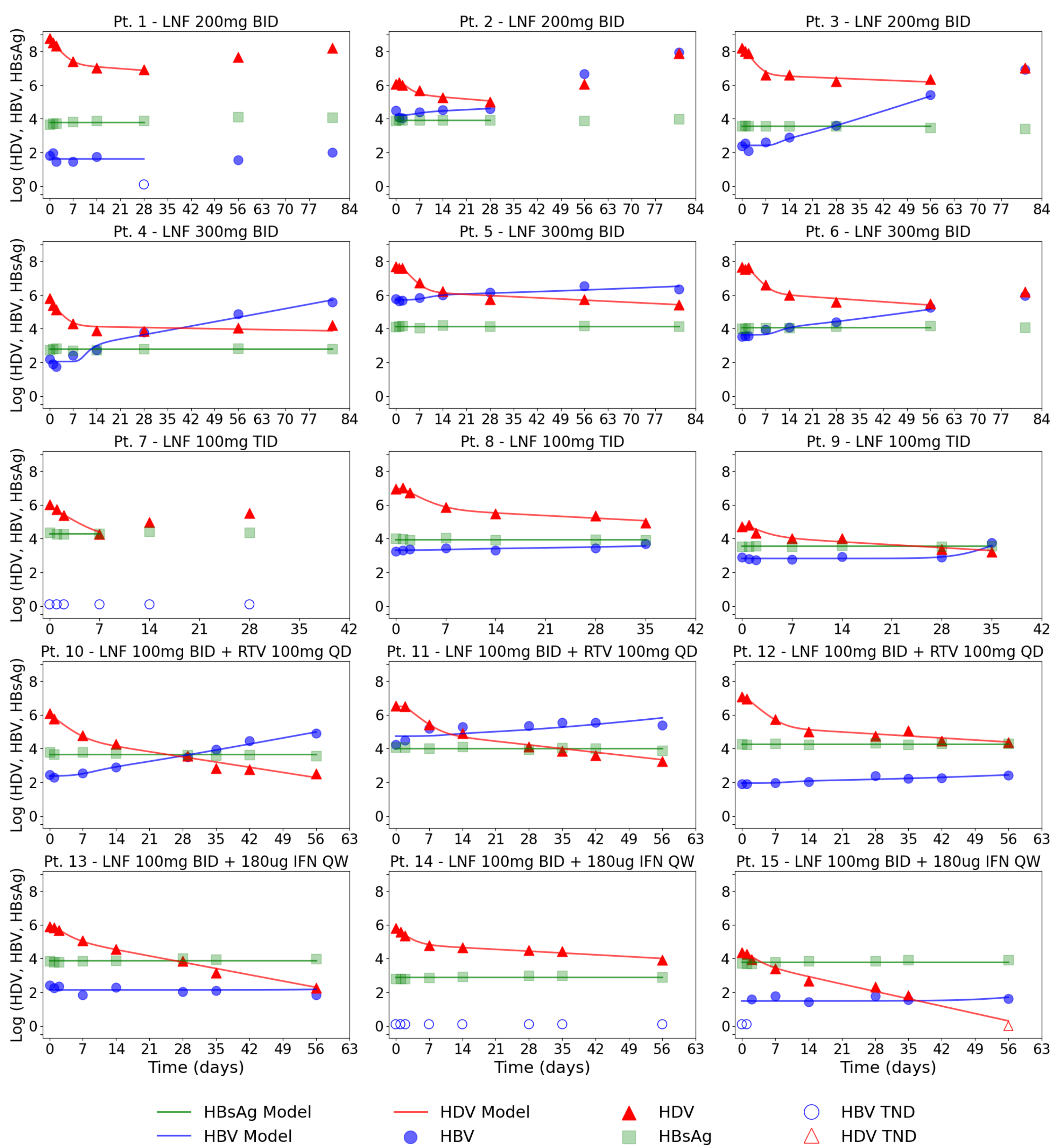

Pt. 1 - LNF 200mg BID
Pt. 2 - LNF 200mg BID
Pt. 3 - LNF 200mg BID
Pt. 4 - LNF 300mg BID
Pt. 5 - LNF 300mg BID
Pt. 6 - LNF 300mg BID
Pt. 7 - LNF 100mg TID
Pt. 8 - LNF 100mg TID
Pt. 9 - LNF 100mg TID
Pt. 10 - LNF 100mg BID + RTV 100mg QD
Pt. 11 - LNF 100mg BID + RTV 100mg QD
Pt. 12 - LNF 100mg BID + RTV 100mg QD
Pt. 13 - LNF 100mg BID + 180ug IFN QW
Pt. 14 - LNF 100mg BID + 180ug IFN QW
Pt. 15 - LNF 100mg BID + 180ug IFN QW
Log (HDV, HBV, HBsAg)
Time (days)
HBsAg Model
HBV Model
HDV Model
HBV
HDV
HBsAg
HBV TND
HDV TND

**Fig. 2**: Modeling (solid lines) and measured serum HDV RNA (triangles), HBV DNA (circles), and HBsAg (squares) in patients undergoing anti-HDV treatment under one of five different regimens. Each row corresponds to one of these groups, with group 1 at the top and descending by row to group 5 [group 1: LNF 200mg twice daily (BID) for 12 weeks; group 2: LNF 300mg BID for 12 weeks; group 3: LNF 100mg thrice daily (TID) for 8 weeks; group 4: LNF 100mg BID + RTV 100mg once daily (QD) for 8 weeks; group 5: LNF 100mg BID + PEG-IFN$\alpha$ 180$\mu$g once daily (QW) for 8 weeks]. Model curves for each patient's data are presented over treatment durations of 84 days (patients 1 to 6), 28 days (patient 7), 35 days (patients 8 and 9), and 56 days (patients 10 to 15). Data points falling below the limit of detection are shown as individual markers only (open circles and triangles) and are not connected by fitting lines, as no curve is fitted to undetectable values**.** TND indicates target not detected. For Patient’s 7- 9 the treatment duration was limited to 35 days due to interruption of the drug supply (Yurdaydin et al. 2018). Patient 7 had a missing data point on day 35.

# Tables

| | | | | | | Baseline | | | Phase 1 | | | Phase 2 | | |
|---|---|---|---|---|---|---|---|---|---|---|---|---|---|---|
| Pt ID | Group | Dosage | VKP | $\tau$ (D) | g | HBV DNA | HDV RNA | HBsAg | Slope (log10/day) | Duration (day) | Magnitude of decline (log10) | Slope (log10/day) | Duration (day) | Magnitude of decline (log10) |
| 1 | 1 | LNF 200mg BID | FPR | 0 | 0.036 | 1.81 | 8.78 | 3.65 | -0.195 | 7 | 1.39 | -0.019 | 21 | 0.46 |
| 2 | 1 | LNF 200mg BID | FPR | 2 | 0.046 | 4.48 | 6.06 | 3.90 | -0.062 | 7 | 0.38 | -0.030 | 21 | 0.68 |
| 3 | 1 | LNF 200mg BID | FPR | 1 | 0.019 | 2.38 | 8.19 | 3.57 | -0.234 | 7 | 1.6 | -0.006 | 49 | 0.24 |
| 4 | 2 | LNF 300mg BID | FPR | 0 | 0.007 | 2.18 | 5.80 | 2.72 | -0.200 | 7 | 1.5 | 0.001 | 77 | 0.1 |
| 5 | 2 | LNF 300mg BID | FPR | 2 | 0.023 | 5.77 | 7.70 | 4.11 | -0.112 | 14 | 1.48 | -0.009 | 70 | 0.8 |
| 6 | 2 | LNF 300mg BID | FPR | 2 | 0.033 | 3.52 | 7.66 | 4.02 | -0.127 | 14 | 1.67 | -0.011 | 42 | 0.5 |
| 7 | 3 | LNF 100mg TID | M | 0 | 0.052 | 0.10 | 6.02 | 4.36 | -0.244 | 7 | 1.74 | N/A | N/A | N/A |
| 8 | 3 | LNF 100mg TID | FPR | 1 | 0.049 | 3.20 | 6.96 | 4.00 | -0.169 | 7 | 1.1 | -0.028 | 28 | 0.93 |
| 9 | 3 | LNF 100mg TID | BP | 1 | 0.056 | 2.90 | 4.73 | 3.54 | -0.107 | 7 | 0.71 | -0.032 | 28 | 0.8 |
| 10 | 4 | LNF 100mg BID + RTV 100mg QD | BP | 0 | 0.102 | 2.45 | 6.08 | 3.79 | -0.128 | 14 | 1.82 | -0.043 | 42 | 1.75 |
| 11 | 4 | LNF 100mg BID + RTV 100mg QD | BP | 2 | 0.075 | 4.23 | 6.54 | 4.07 | -0.169 | 7 | 1.11 | -0.045 | 49 | 2.18 |
| 12 | 4 | LNF 100mg BID + RTV 100mg QD | FPR | 1 | 0.040 | 1.91 | 7.07 | 4.28 | -0.153 | 14 | 2.07 | -0.016 | 42 | 0.64 |
| 13 | 5 | LNF 100mg BID + 180ug IFN QW | BP | 1 | 0.122 | 2.40 | 5.91 | 3.84 | -0.124 | 7 | 0.86 | -0.058 | 49 | 2.79 |
| 14 | 5 | LNF 100mg BID + 180ug IFN QW | BP | 0 | 0.035 | 0.10 | 5.79 | 2.80 | -0.136 | 7 | 1 | -0.017 | 49 | 0.87 |
| 15 | 5 | LNF 100mg BID + 180ug IFN QW | BP | 1 | 0.142 | 0.10 | 4.37 | 3.73 | -0.120 | 14 | 1.69 | -0.066 | 42 | 2.68 |
| Median (IQR) | | | | | 0.046 [0.034,0.07]] | 2.40 [1.86-3.36] | 6.08 [5.86-7.37] | 3.84 [3.61-4.05] | -0.14 [-0.18, -0.12] | 7 [7, 14] | 1.48 [1.05, 1.68] | -0.024 [-0.04, -0.012] | 42 [32, 49] | 0.8 [0.54, 1.54] |

**Table 1**. Summary of HDV Kinetic Patterns. VKP= HDV RNA Kinetic Pattern; BP= Biphasic; FPR= Flat Partial Response; M=Monophasic; D=days; IQR=Interquartile range; IFN, pegylated interferon-α

| | Population estimates | | Inter-individual variability | |
|---|---|---|---|---|
| Parameter (Units) | Value [95% Confidence interval] | RSE% | Value [95% Confidence interval] | RSE% |
| $D_0$ $\log_{10}$(IU/mL) | 6.35 [5.83 – 6.82] | 4.81 | 0.19 [0.12 – 0.22] | 18.4 |
| $\varepsilon$ (%) | 0.94 [0.89 – 0.97] | 1.74 | 0.99 [0.38 – 1.27] | 24.9 |
| c (day$^{-1}$) | 0.55 [0.50 – 0.68] | 11.0 | 0.08 [0.05 – 0.27] | 74.8 |
| g | 0.043 [0.026 – 0.069] | 23.4 | 0.77 [0.50 – 0.94] | 21.3 |
| $B_0$ $\log_{10}$(IU/mL) | 2.33 [1.77 – 2.96] | 14.9 | 0.57 [0.36 – 0.75] | 20.5 |
| $\kappa$ $\log_{10}$(IU/mL) | 2.78 [0.99 – 4.77] | 33.6 | 0.79 [0.18 – 1.63] | 38.1 |
| n | 1.04 [1.00 – 2.91] | 8.39 | 4.39 [1.28 – 8.53] | 47.3 |
| $H_0$ $\log_{10}$(IU/mL) | 3.75 [3.54 – 3.93] | 3.18 | 0.12 [0.06 – 0.16] | 18.4 |

**Table 2**: Estimates of the model parameters are presented along with their relative standard errors (RSE%) and 95% confidence intervals. In NLME models the population estimate is described by the fixed effects and the inter-individual variability by the standard deviation (SD) of the random effects. The fixed effect estimates represent the average effect of the parameter across the population being studied, while the random effect represent the variation in the effect of the parameter across individuals within the population. The RSE% values provide an indication of the uncertainty or variability associated with each estimate. In general, (RSE < 50%) indicate greater precision in the estimate, while higher values indicate greater uncertainty. For parameter n, we used bootstrapping to compute its confidence interval.

| Gr | Patient | Dose | F/Time (days) | N | $\tau$ days | $D_0$ $(log_{10}IU/ml)$ | $\varepsilon$ % | c $day^{-1}$ | g | $B_0$ $(log_{10}IU/ml)$ | $\kappa$ $(log_{10}IU/ml)$ | n | $H_0$ $(log_{10}IU/ml)$ |
|---|---|---|---|---|---|---|---|---|---|---|---|---|---|
| | 1 | LNF 200mg BID | 28/84 | 6 | 0 | 8.75 | 0.97 | 0.55 | 0.036 | 1.61 | 2.78 | 1.04 | 3.78 |
| | 2 | | 28/84 | 6 | 2 | 6.11 | 0.77 | 0.54 | 0.046 | 4.29 | 5.17 | 1.03 | 3.92 |
| **1** | 3 | | 56/84 | 7 | 1 | 8.04 | 0.96 | 0.57 | 0.019 | 2.42 | 6.57 | 7.52 | 3.56 |
| **2** | 4 | LNF 300mg BID | 84/84 | 6 | 0 | 5.66 | 0.97 | 0.55 | 0.007 | 2.06 | 4.19 | 11.89 | 2.79 |
| | 5 | | 84/84 | 8 | 2 | 7.63 | 0.96 | 0.54 | 0.023 | 5.72 | 6.17 | 1.02 | 4.14 |
| | 6 | | 56/84 | 7 | 2 | 7.59 | 0.97 | 0.55 | 0.033 | 3.63 | 6.22 | 1.88 | 4.07 |
| **3** | 7 | LNF 100mg TID | 7/35 | 4 | 0 | 5.90 | 0.98 | 0.57 | 0.052 | 0.97 | 2.74 | 1.04 | 4.30 |
| | 8 | | 35/35 | 7 | 1 | 6.97 | 0.94 | 0.55 | 0.049 | 3.40 | 2.95 | 1.04 | 3.96 |
| | 9 | | 35/35 | 7 | 1 | 4.74 | 0.80 | 0.55 | 0.056 | 2.82 | 3.41 | 7.60 | 3.55 |
| **4** | 10 | LNF 100mg BID + RTV 100mg QD | 56/56 | 8 | 0 | 6.05 | 0.96 | 0.54 | 0.102 | 2.37 | 4.59 | 1.17 | 3.67 |
| | 11 | | 56/56 | 9 | 2 | 6.49 | 0.97 | 0.55 | 0.075 | 4.76 | 4.39 | 1.01 | 4.02 |
| | 12 | | 56/56 | 5 | 1 | 6.98 | 0.98 | 0.55 | 0.040 | 1.95 | 4.73 | 1.03 | 4.27 |
| **5** | 13 | LNF 100mg BID+ 180ug IFN QW | 56/56 | 8 | 1 | 5.90 | 0.86 | 0.55 | 0.122 | 2.13 | 1.48 | 1.11 | 3.88 |
| | 14 | | 56/56 | 8 | 0 | 5.79 | 0.89 | 0.55 | 0.035 | 0.89 | 2.63 | 1.04 | 2.88 |
| | 15 | | 56/56 | 8 | 1 | 4.33 | 0.84 | 0.55 | 0.142 | 1.45 | 0.77 | 1.02 | 3.80 |
| **(1-5)** | **Median [IQR]** | | | | | **6.10 [5.82, 7.44]** | **0.96 [0.87, 0.97]** | **0.55 [0.547, 0.574]** | **0.046 [0.034, 0.07]** | **2.37 [1.70, 3.57]** | **4.19 [2.75, 5.06]** | **1.04 [1.03, 1.70]** | **3.88 [3.59, 4.05]** |

**Table 3**: Best model fit results for different treatment regimens involving 15 patients, along with their confidence intervals. The Gr column indicates individual groups, while the Patient column indicates the patient number. The Dose column shows the dosage taken by each group, and the F/Time column displays the number of days considered for fittings per patient over the total duration of therapy, respectively. N represents the total number of time points for each patient. IQR refers to the interquartile range. The pharmacological delay in (days) is presented without confidence intervals because it was estimated directly from the kinetic data without employing any modeling techniques. The error model parameters estimated as 0.24 [95% CI : 0.18 – 0.31], 0.06 [0.05 – 0.07], 0.17 [0.14 – 0.20] for B, H, and D respectively.

## Supplementary Materials

<u>**A: LNF and ALT Kinetic analysis**</u>

### 4.1.1 LNF Kinetics

Here, we present LNF pharmacokinetic analysis in a limited capacity as this is the not the main focus of the manuscript.

Over the first 2 days of treatment, all patients experienced a rapid increase in LNF concentrations following the first dose. After this initial increase, LNF concentration values remained within 1$log_{10}$ in the majority of patients (**Fig. S1**). No major differences were apparent in groups receiving LNF monotherapy at different doses (**Table S4**). Of interest, $C_{last}$ tended to be slightly higher in LNF+RTV group compared to other groups, although this difference was not statistically significant (p=0.18). The detailed pharmacokinetic analysis was presented in **Table S4.**

For 13/15 patients, the slope of LNF concentration did not differ significantly from zero (linear regression, p<0.05), plateauing at a median value of 3.14 $log_{10}$ mg (IQR 0.30) until the EOT (n=9) or beginning of HDV viral breakthrough (n=4). There were no differences in LNF plateau value across dosage groups (p=0.182).

One patient (patient 7) had a slight decline in LNF during which a decline in HDV viral load was also seen, as previously reported (Yurdaydin et al. 2018). This patient experienced a viral breakthrough after day 7 and stopped treatment early at day 28. A decrease in LNF serum levels was seen in accordance with the beginning of the HDV viral rebound (**Fig. S1**). Another patient (Patient 12) had a slight increase in LNF serum values during treatment, but only by <1 $log_{10}$ (**Fig. S1**).

Patient 5 is an interesting case because while LNF concentration increased at the beginning of therapy (like the rest of the patients), the concentration then decreased from day 2 until day 30 and then plateaued at a median of 2.8 $log_{10}$ mg (IQR 1.2) (**Fig. S1**).

#### 4.1.2 ALT Kinetics

The median baseline ALT was 82 (IQR: 109) U/L. Six patients had a high baseline ALT level, (> 100 U/L) and the remaining 9 patients had baseline ALT levels ranging from 23-84 U/L (**Table S1**, **Fig. S1**). There were no associations between baseline ALT values and treatment group, treatment type (monotherapy or combination therapy), or HDV viral kinetic pattern.

For ALT kinetics, a variety of patterns were observed. Six patients had a high baseline ALT with a median of 106.5 U/L (IQR 75.8), which declined to a median value of 62 U/L (IQR) at the end of treatment (EOT) or before viral rebound. Five of the 9 remaining patients experienced a plateau in ALT values throughout treatment, 3 experienced a slight decline, and 1 a slight increase (**Fig. S1**).

The slope of ALT decline during treatment was not associated with viral kinetic pattern (p=0.067), LNF dosage group (p=0.212), combination vs monotherapy (p=0.088), or IFN or non-IFN based treatments (p=0.101).

**B: Monolix parameter estimation and Statistical model**

Using the individual fits, we were able to estimate the parameter values for each patient, capturing unique characteristics. However, this approach might not fully represent the variability and patterns across the entire patient population.

To analyze the longitudinal data, we thus follow a population approach using a nonlinear mixed effects (NLME) model. In this model for each individual i (i = 1, …, N), the $j^{th}$ log-transformed observation (j = 1,…,$N_i$), in the $k^{th}$ viral bio-marker (namely HDV RNA, HBV DNA, HBsAg) at time $t_{ijk}$ was

$$Y_{ijk} = f\left(\theta_i, t_{ijk}\right) + \ e_{ijk} \qquad \text{(Eq. S1)}$$

Here, $\theta_i$ represents the vector of parameters for individual i, while $e_{ijk}$ represents the residual error of patient i at observation j for each marker k, assumed to follow a normal distribution with mean zero and variance $\sigma^2$. Each $\theta_i$ comes from a population distribution described by the fixed effects vector $\mu$, the vector of random effects $\eta_i$ describing the inter-individual variability with variance-covariance matrix $\Omega^2$, and the vector of individual baseline covariates $z_i$. For the parameter $\varepsilon$, which represents the efficacy of the treatment, and the parameter (n), that governs the rate increase in HBV production under anti-HDV treatment, where $0 < \varepsilon < 1$ and $1 < n$, we assumed a logit-normal distribution as shown in (**Eq. S2**). All the other individual parameters were assumed to be log-normally distributed (**Eq. S3**):

$$\theta_i = \frac{\frac{\mu}{1+\mu}}{\frac{\mu}{1+\mu}+e^{-(\eta_i+z_i)}} \qquad \text{(Eq.S2)}$$

$$\theta_i = \mu e^{(\eta_i+z_i)} \qquad \text{(Eq. S3)}$$

The parameter estimation was performed by maximizing the likelihood using the stochastic approximation expectation maximization (SAEM) algorithm implemented in Monolix software (www.lixoft.com, v2023R1, Orsay, France). The log-likelihood was estimated using the importance sampling approach, which is based on reweighting the simulated points to increase the efficiency and accuracy of parameter estimation. The Fisher Information Matrix (FIM) was calculated using a stochastic approximation method. We censored data below the detection limit, using 20 IU/mL for HBV and 50 IU/mL for HDV as the lower limits of detection (LLoD), as described in Section 2.3. The maximum likelihood was computed by the information provided by the censored and non-censored data. The parameter $\tau$ was not identifiable, and to improve the results of our model, $\tau$ was incorporated as a regressor. The parameter $c_H$ was fixed to 0.53 day$^{-1}$ across all individuals, hence it was also incorporated as a regressor. We employed a constant error model to understand the discrepancies between observed data and our model predictions for HBV, HDV, and HBsAg.

Confidence intervals were computed at the 95% level of significance using the bootstrap method. Confidence intervals for all the population parameter estimates are presented in **Table 3**.

We used R version 4.3.2 to compute the confidence intervals for our population parameter estimates, using data linked with Monolix.

## C: Model with both infectious and non-infectious virus

Verrier et al. reported that in vitro LNF treatment in chronic HDV increases the proportion of non-infectious viral particles by promoting the accumulation of edited HDV genomes, leading to reduced overall infectivity (Verrier et al. 2022). To account for this potential mechanism in human data, we extended the model presented in the main manuscript (**Eq. 1**) to include both infectious and non-infectious virus. The new model is detailed in (**Eq. S4**) and illustrated in **Fig. S2**. It is described by the following equations:

$$\frac{dD_i}{dt} = (1-\epsilon)p_i I_0 \, e^{-\mu t} e^{-g(t-\tau)} - cD_i$$

$$\frac{dD_n}{dt} = (1-\epsilon)p_i I_0 (1- \, e^{-\mu t})e^{-g(t-\tau)} + \, (1-\epsilon)p_n I_0 e^{-g(t-\tau)} - cD_n \qquad \text{(Eq. S4)}$$

$$\frac{dB}{dt} = p_b I_0 \, (1 + \left(\frac{\kappa}{D_i+D_n+LoD}\right)^n) - cB$$

$$\frac{dH}{dt} = p_h I_0 - c_h H$$

where $D_i$, $D_n$, and B represent the infectious HDV RNA, non-infectious HDV RNA and the HBV DNA respectively. H denotes HBsAg. Parameters $p_i$, $p_n$, $p_b$, and $p_h$ represent the steady state production rates of HDV RNA (infectious), HDV RNA (non-infectious), HBV DNA and HBsAg, respectively, from the infected cell number $I_0$. We assumed that the clearance rate constant, *c*, is the same for HDV RNA and HBV DNA and that it is within the range of the HDV RNA clearance rate estimated in (Koh et al. 2015; Shekhtman et al. 2020). Parameter ε represents the efficacy in blocking the production rate of HDV RNA, both infectious and non-infectious, with $0 < \epsilon < 1$, and *g* is the assumed additional treatment inhibitory effect in blocking HDV RNA production, reminiscent of previous modeling efforts of antiviral treatment during chronic HBV mono-infection in patients and humanized mice (Kadelka, Dahari, and Ciupe 2021; Reinharz et al. 2021). The transition from infectious to non-infectious HDV virus over time under LNF treatment is modeled by exp(-$\mu$t), where $\mu$ is a constant, and *t* is the time post treatment initiation. Parameter *n* governs the HBV DNA production rate increase under anti-HDV treatment. Parameter $\kappa$ represents the threshold total serum HDV RNA level that triggers an increase in HBV DNA production. LoD is the lower limit of detection of the HDV RNA assay.

Before treatment, HDV RNA (infectious), HDV RNA (non-infectious), HBV DNA and HBsAg levels are in equilibrium with the production rate of HDV RNA (infectious) given by

$$p_i = \frac{cD_{i0}}{I_0}$$

that of HDV RNA (non-infectious) represented by

$$p_n = \frac{cD_{n0}}{I_0}$$

that of HBV DNA represented by

$$p_b = \frac{cD_{b0}}{I_0\left(1 + \left(\frac{\kappa}{(D_{i0} + D_{n0} + LoD)}\right)^n\right)}$$

and that of HBsAg represented by

$$p_h = \frac{cH_0}{I_0}$$

The number of the infected cells, $I_0$, was kept constant under treatment and prior to treatment onset.

Parameter estimation

We fitted the total HDV RNA for patient 10 using Berkeley Madonna version 10.4.2, Total HDV RNA = Infectious HDV RNA + Non-infectious HDV RNA. We assumed the number of co-infected cells ($I_0$) to be a constant at $1 \times 10^6$. We varied the values of $\mu$ with $\mu$ = 0.001, $\mu$=0.01 and $\mu$=0.1, and presented a sensitivity analysis in **Fig. S3**.

Results

The extended model  (**Eq. S4**, **Fig. S2**) accounts for the change in the ratio of infectious vs non-infectious particles as was done previously for HCV (Goyal et al. 2017). Due to lack of knowledge about the ratio of infectious to non-infectious virions, we chose to explore three scenarios (infectious vs non-infectious; 90% vs 10%, 50% vs 50%, and 10% vs 90%) (**Fig. S3**). For illustrative purposes, we present application of the extended model to a representative case( patient 10) 10, for whom no delay ($\tau = 0$) between treatment initiation and the onset of measurable antiviral effect on HDV RNA was observed. Regardless of the chosen ratio, using data from patient 10 as a basis, there was little effect on the parameter estimates for $D_0$, $\epsilon$, g, and c, and they remained very close to those estimated by the primary model (**Eq. 1**). As $\mu$ increases, the model predicts a progressively faster decline in infectious HDV across all simulated scenarios (Fig. S3), reflecting an accelerated transition from infectious to non-infectious viral particles under LNF based therapy. This effect is consistently observed regardless of the assumed ratio of infectious to non-infectious HDV RNA. However, despite differences in the infectious HDV RNA curves for higher $\mu$ values (Fig. S3 c,f,i) compared with lower $\mu$ values (Fig S3 a,d,g), total HDV RNA remains largely unchanged across all panels. Since the total HDV RNA is measured in the clinical data, these internal shifts in viral composition cannot be identified, and the estimated parameters remain close to those obtained with the primary model (**Eq. 1**). Therefore, extending the model to explicitly include infectious and non-infectious HDV RNA compartments appears premature in the absence of data on their relative proportions.

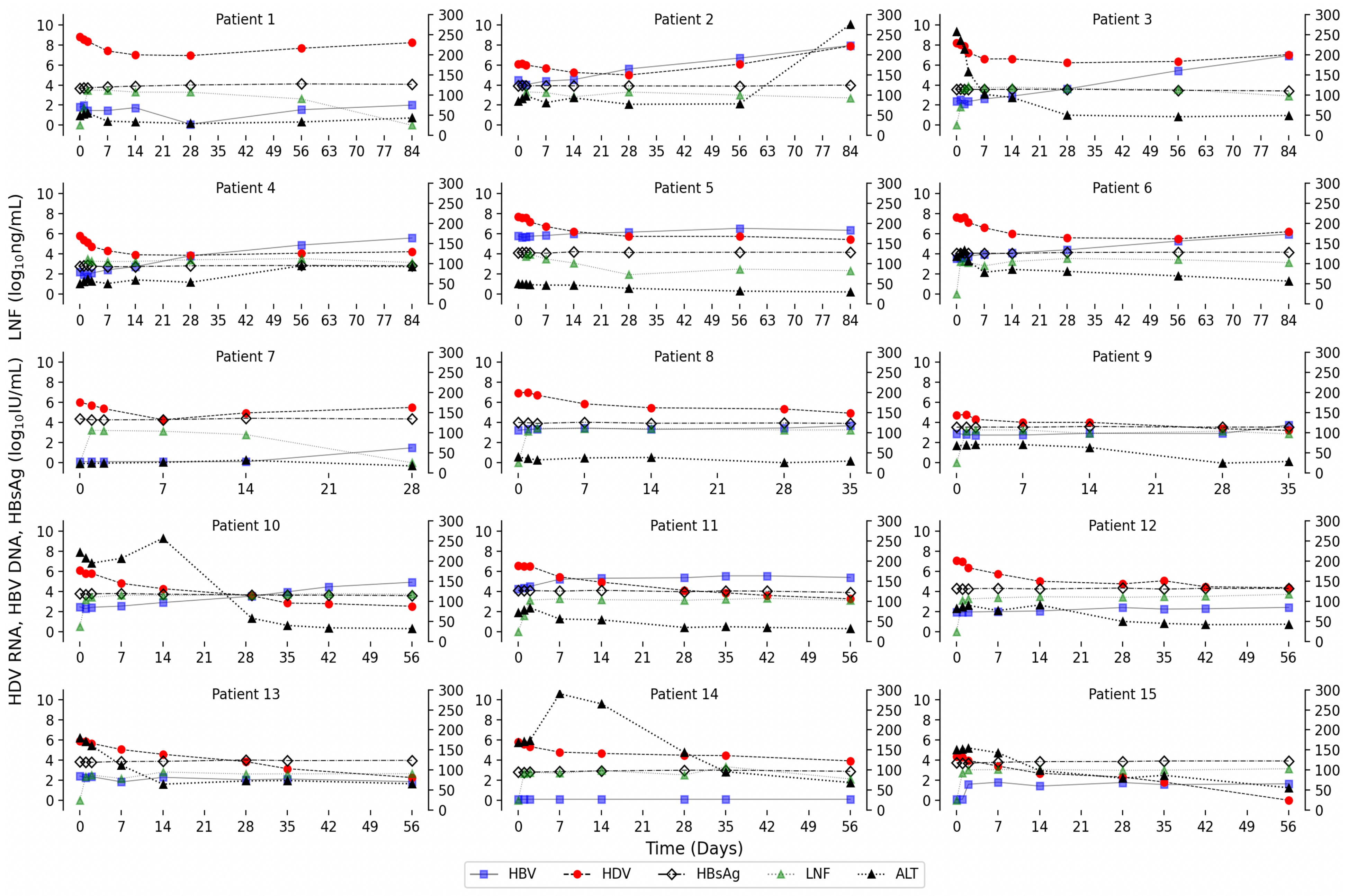

**Fig. S1:** HDV RNA (filled red circles), HBV DNA (filled blue boxes), HBsAg (empty diamonds), LNF concentration (filled green triangles), and ALT (filled black triangles) kinetics during LNF based therapy.

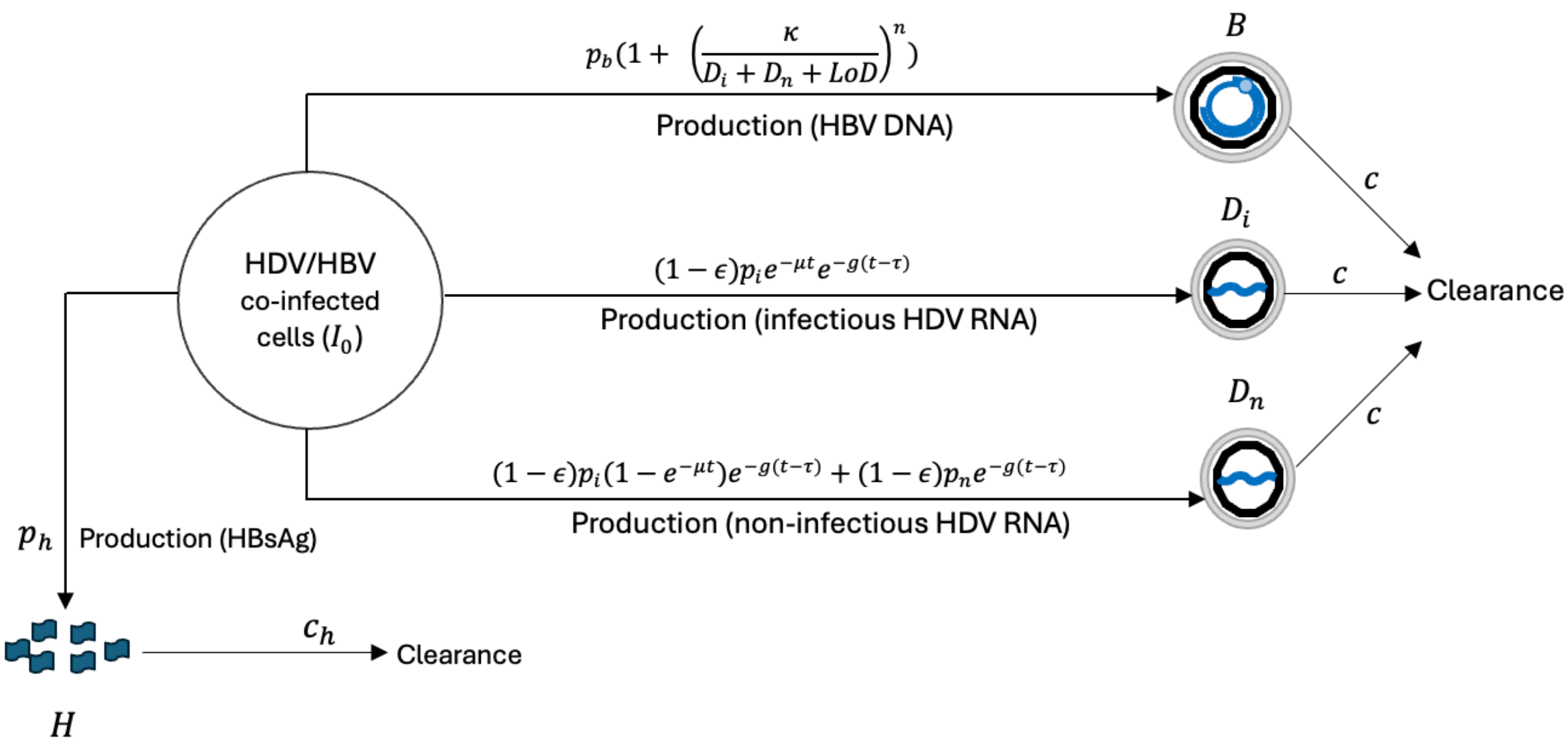

**Fig. S2**: Schematic diagram for our model incorporating non-infectious and infectious virus (**Eq. S4**). In the Figure the co-infected cells ($I_0$) produce HDV infectious virions ($D_i$), HDV non-infectious virions ($D_n$), HBV virions ($B$) and HBsAg ($H$). The production rates for $D_i$, $D_n$, $B$, and $H$ are denoted by $p_i$, $p_n$, $p_b$, and $p_h$ respectively. $H$ is cleared at rate $c_h$, while for $D_i$, $D_n$, $B$ are cleared at the rate $c$. The parameter g denotes the continous decline in HDV RNA production, with $\tau$ being the pharmacological time delay. The expression exp(-$\mu$ $t$) denotes a gradual increase in the ratio of non-infectious to infectious virus, which is modelled by exp (-$\mu$ t), where $\mu$ is a constant and t represents time post treatment initiation. Parameter $\varepsilon$, denotes the treatment efficacy in blocking viral production ($0 < \varepsilon < 1$). Parameter *n* governs the HBV production rate increase under anti-HDV treatment, while parameter $\kappa$ represents the threshold total HDV RNA level in blood that triggers an increase in HBV DNA production.

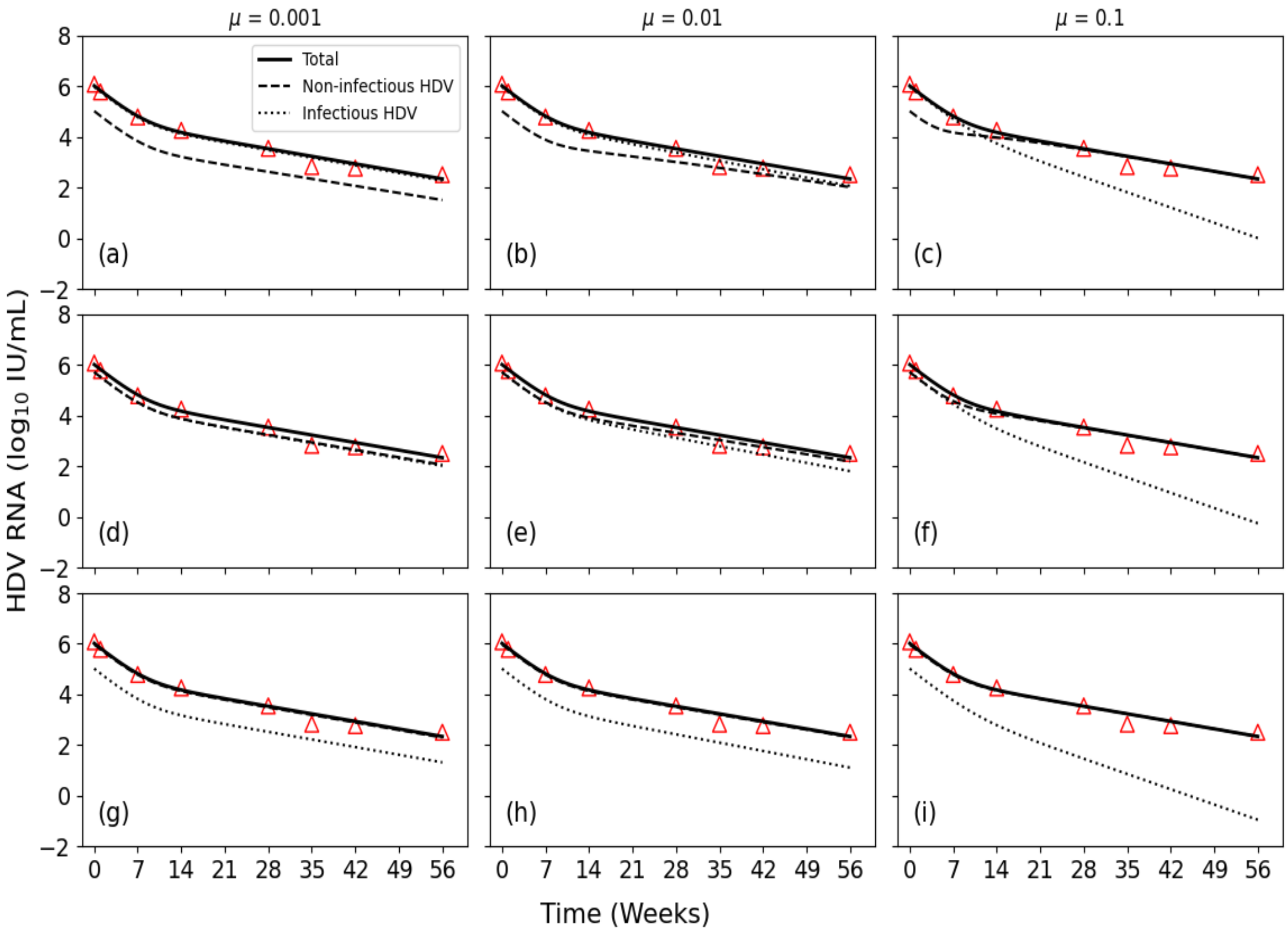

**Fig. S3**: Sensitivity analysis of parameter $\mu$ in (**Eq. S4**), on HDV RNA only. This parameter represents the rate at which the ratio of non-infectious to infectious HDV RNA increases over time due to LNF treatment. Panels (a-c) show a scenario with 90% infectious HDV RNA and 10% non-infectious HDV RNA, panels (d-f) show a 50% infectious HDV RNA and 50% non-infectious HDV RNA scenario, and panels (g-i) depict a scenario with 10% infectious HDV RNA and 90% non-infectious HDV RNA. Panels (a, d, g) in the left column correspond to $\mu$ = 0.001, panels (b, e, h) in the center column correspond to $\mu$ = 0.01, and panels (c, f, i) in the right column correspond to $\mu$ = 0.1. All other parameters were set at ($D_0$=6.05, g = 0.101, c = 0.54, $\epsilon = 0.96$). The red triangles denote the measured HDV data.

**Tables**

| Group | Dose | Pt | Sex | Age | HDV RNA | HBV DNA | HBsAg | ALT | Prev. IFN |
|---|---|---|---|---|---|---|---|---|---|
| 1 | LNF 200mg BID N=3 | 1 | F | 57 | 6.91 | 1.81 | 3.65 | 49 | 2/3 |
| | | 2 | M | 63 | 4.20 | 4.48 | 3.90 | 84 | |
| | | 3 | F | 58 | 6.44 | 2.38 | 3.87 | 258 | |
| | | | | **59** | **5.85** | **2.89** | **3.81** | **130** | |
| 2 | LNF 300mg BID N=3 | 4 | M | 36 | 3.94 | 2.18 | 2.75 | 50 | 2/3 |
| | | 5 | M | 38 | 5.84 | 5.77 | 4.10 | 50 | |
| | | 6 | F | 46 | 5.80 | 3.52 | 4.02 | 119 | |
| | | | | **40** | **5.19** | **3.82** | **3.62** | **73** | |
| 3 | LNF 100mg TID N=3 | 7 | F | 40 | 6.02 | 0 | 4.36 | 23 | 2/3 |
| | | 8 | F | 48 | 6.96 | 3.23 | 4.00 | 39 | |
| | | 9 | M | 51 | 4.73 | 2.90 | 3.54 | 67 | |
| | | | | **46** | **5.90** | **2.04** | **3.97** | **43** | |
| 4 | LNF 100mg BID + RTV 100mg QD N=3 | 10 | M | 43 | 6.08 | 2.45 | 3.79 | 222 | 2/3 |
| | | 11 | M | 49 | 6.54 | 4.23 | 4.07 | 72 | |
| | | 12 | M | 61 | 7.07 | 1.91 | 4.28 | 82 | |
| | | | | **51** | **6.56** | **2.86** | **4.05** | **125** | |
| 5 | LNF 200mg BID + PEG-IFN$\alpha$ 180$\mu$g QW N=3 | 13 | M | 22 | 5.91 | 2.40 | 3.84 | 180 | 2/3 |
| | | 14 | F | 52 | 5.79 | 0 | 2.80 | 168 | |
| | | 15 | M | 49 | 4.37 | 0 | 3.73 | 150 | |
| | | | | **41** | **5.36** | **0.8** | **3.46** | **166** | |
| ALL | N=15 | | 9 Males/ 6 Females | 49 (22-63) | 5.91 (3.94 – 7.07) | 2.40 (0.00 – 5.77) | 3.87 (2.75 – 4.36) | 82 (23 - 258) | 14/20 |

**Table S1**: Patient Characteristics at time of enrollment. Mean values for each group, as well as individual patient values, are presented. For the All-patient row, data are given as median and range.

| Group | Patient number | Dose | Point of rebound | Rebound ($log_{10}$ IU/mL) |
|---|---|---|---|---|
| 1 | 1 | LNF 200mg BID | day 28 to day 56 | 0.73 |
| 1 | 2 | LNF 200mg BID | day 28 to day 56 | 1.07 |
| 1 | 3 | LNF 200mg BID | day 56 to day 84 | 0.67 |
| 2 | 6 | LNF 300mg BID | day 56 to day 84 | 0.71 |
| 3 | 7 | LNF 100mg TID | day 7 to day 14 | 0.68 |

**Table S2:** Summarized values for breakthroughs defined as an increase of HDV RNA by $\geq 0.5$ $log_{10}$ from the previous measurement observed during treatment. Notably, these rebounds were only seen in patients receiving LNF monotherapy.

| | | | | | Baseline | | | Phase 1 | | | Phase 2 | | |
|---|---|---|---|---|---|---|---|---|---|---|---|---|---|
| Pt ID | Dosage | VKP | $\tau$ (D) | g | HBV DNA | HDV RNA | HBsAg | Slope (log10/day) | Duration (day) | Magnitude of decline (log10) | Slope (log10/day) | Duration (day) | Magnitude of decline (log10) |
| 1 | LNF 200mg BID | FPR | 0 | 0.036 | 1.81 | 8.78 | 3.65 | -0.195 | 7 | 1.39 | -0.019 | 21 | 0.46 |
| 2 | LNF 200mg BID | FPR | 2 | 0.046 | 4.48 | 6.06 | 3.90 | -0.062 | 7 | 0.38 | -0.030 | 21 | 0.68 |
| 3 | LNF 200mg BID | FPR | 1 | 0.019 | 2.38 | 8.19 | 3.57 | -0.234 | 7 | 1.6 | -0.006 | 49 | 0.24 |
| median<br>[minimum, maximum] | | | | 0.036<br>[0.019,0.046] | 2.38<br>[1.81,4.48] | 8.19<br>[6.06,8.78] | 3.65<br>[3.57,3.90] | -0.195<br>[-0.234,-0.062] | 7<br>[7,7] | 1.39<br>[0.38,1.6] | -0.019<br>[-0.006,-0.03] | 21<br>[21,49] | 0.46<br>[0.24,0.68] |
| 4 | LNF 300mg BID | FPR | 0 | 0.007 | 2.18 | 5.80 | 2.72 | -0.200 | 7 | 1.5 | 0.001 | 77 | 0.1 |
| 5 | LNF 300mg BID | FPR | 2 | 0.023 | 5.77 | 7.70 | 4.11 | -0.112 | 14 | 1.48 | -0.009 | 70 | 0.8 |
| 6 | LNF 300mg BID | FPR | 2 | 0.033 | 3.52 | 7.66 | 4.02 | -0.127 | 14 | 1.67 | -0.011 | 42 | 0.5 |
| median<br>[minimum, maximum] | | | | 0.023<br>[0.007,0.033] | 3.52<br>[2.18,5.77] | 7.66<br>[5.8,7.7] | 4.02<br>[2.72,4.11] | -0.127<br>[-0.2,-0.112] | 14<br>[7,14] | 1.5<br>[1.48,1.67] | -0.009<br>[-0.001,-0.011] | 70<br>[42,77] | 0.5<br>[0.1,0.8] |
| 7 | LNF 100mg TID | M | 0 | 0.052 | 0.10 | 6.02 | 4.36 | -0.244 | 7 | 1.74 | N/A | N/A | N/A |
| 8 | LNF 100mg TID | FPR | 1 | 0.049 | 3.20 | 6.96 | 4.00 | -0.169 | 7 | 1.1 | -0.028 | 28 | 0.93 |
| 9 | LNF 100mg TID | BP | 1 | 0.056 | 2.90 | 4.73 | 3.54 | -0.107 | 7 | 0.71 | -0.032 | 28 | 0.8 |
| median<br>[minimum, maximum] | | | | 0.052<br>[0.049,0.056] | 2.9<br>[0.1,3.2] | 6.02<br>[4.73,6.96] | 4<br>[3.54,4.36] | -0.169<br>[-0.107,-0.244] | 7<br>[7,7] | 1.1<br>[0.71,1.74] | -0.03<br>[-0.028,-0.032] | 28<br>[28,28] | 0.87<br>[0.8,0.93] |
| 10 | LNF 100mg BID + RTV 100mg QD | BP | 0 | 0.102 | 2.45 | 6.08 | 3.79 | -0.128 | 14 | 1.82 | -0.043 | 42 | 1.75 |
| 11 | LNF 100mg BID + RTV 100mg QD | BP | 2 | 0.075 | 4.23 | 6.54 | 4.07 | -0.169 | 7 | 1.11 | -0.045 | 49 | 2.18 |
| 12 | LNF 100mg BID + RTV 100mg QD | FPR | 1 | 0.040 | 1.91 | 7.07 | 4.28 | -0.153 | 14 | 2.07 | -0.016 | 42 | 0.64 |
| median<br>[minimum, maximum] | | | | 0.075<br>[0.040,0.102] | 2.45<br>[1.91,4.23] | 6.54<br>[6.08,7.07] | 4.07<br>[3.79,4.28] | -0.153<br>[-0.169, -0.128] | 14<br>[7,14] | 1.82<br>[1.11,2.07] | -0.043<br>[-0.045, -0.016] | 42<br>[42,49] | 1.75<br>[0.64,2.18] |
| 13 | LNF 100mg BID + 180ug PEG-IFN$\alpha$ QW | BP | 1 | 0.122 | 2.40 | 5.91 | 3.84 | -0.124 | 7 | 0.86 | -0.058 | 49 | 2.79 |
| 14 | LNF 100mg BID + 180ug PEG-IFN$\alpha$ QW | BP | 0 | 0.035 | 0.10 | 5.79 | 2.80 | -0.136 | 7 | 1 | -0.017 | 49 | 0.87 |
| 15 | LNF 100mg BID + 180ug PEG-IFN$\alpha$ QW | BP | 1 | 0.142 | 0.10 | 4.37 | 3.73 | -0.120 | 14 | 1.69 | -0.066 | 42 | 2.68 |
| median<br>[minimum, maximum] | | | | 0.122<br>[0.035,0.142] | 0.10<br>[0.10,2.40] | 5.79<br>[4.37,5.91] | 3.73<br>[2.80,3.84] | -0.124<br>[-0.136, -0.12] | 7<br>[7,14] | 1<br>[0.86,1.69] | -0.058<br>[-0.066, -0.017] | 49<br>[42,49] | 2.79<br>[0.87,2.68] |

**Table S3**. Summary of HDV Kinetic Patterns. VKP= HDV RNA Kinetic Pattern; BP= Biphasic; FPR= Flat Partial Response; M=Monophasic; D=days; IQR=Interquartile range

| Patient ID | Dosage | $C_{max}$ log(ng/mL) | $T_{max}$(days) | $C_{last}$ log(ng/mL) | $T_{last}$(days) | $C_{avg}$ log(ng/mL) | $C_{med}$ log(ng/mL) |
|---|---|---|---|---|---|---|---|
| 1 | LNF 200mg BID | 3.48 | 2 | 3.28 | 56 | 3.22 | 3.29 |
| 2 | LNF 200mg BID | 3.31 | 2 | 3.30 | 84 | 3.04 | 3.24 |
| 3 | LNF 200mg BID | 3.85 | 3 | 3.55 | 84 | 3.59 | 3.69 |
| Median [minimum, maximum] | | 3.48 [3.31,3.85] | 2 [2,3] | 2.71 [2.62,2.91] | 84 [56,84] | 3.22 [3.04,3.59] | 3.29 [3.24,3.69] |
| 4 | LNF 300mg BID | 3.51 | 56 | 3.51 | 84 | 3.32 | 3.33 |
| 5 | LNF 300mg BID | 3.96 | 3 | 2.32 | 84 | 3.00 | 3.08 |
| 6 | LNF 300mg BID | 3.74 | 3 | 3.40 | 84 | 3.43 | 3.47 |
| Median [minimum, maximum] | | 3.74 [3.51,3.96] | 3 [3,56] | 3.40 [2.32,3.51] | 84 [84,84] | 3.32 [3.00,3.43] | 3.33 [3.08,3.47] |
| 7 | LNF 100mg TID | 3.22 | 1 | 3.14 | 14 | 3.09 | 3.21 |
| 8 | LNF 100mg TID | 3.44 | 2 | 3.27 | 35 | 3.32 | 3.33 |
| 9 | LNF 100mg TID | 3.26 | 7 | 2.87 | 35 | 3.09 | 3.13 |
| Median [minimum, maximum] | | 3.26 [3.22,3.44] | 2 [1,7] | 3.14 [2.87,3.27] | 35 [14,35] | 3.09 [3.09,3.32] | 3.21 [3.13,3.33] |
| 10 | LNF 100mg BID + RTV 100mg QD | 3.76 | 42 | 3.70 | 56 | 3.14 | 3.63 |
| 11 | LNF 100mg BID + RTV 100mg QD | 3.29 | 42 | 3.11 | 56 | 3.17 | 3.14 |
| 12 | LNF 100mg BID + RTV 100mg QD | 3.70 | 56 | 3.70 | 56 | 3.42 | 3.43 |
| Median [minimum, maximum] | | 3.70 [3.29,3.76] | 42 [42,56] | 3.70 [3.11,3.70] | 56 [56,56] | 3.17 [3.14,3.42] | 3.43 [3.14,3.63] |
| 13 | LNF 100mg BID + 180ug PEG-IFN$\alpha$ QW | 2.81 | 14 | 2.64 | 56 | 2.52 | 2.56 |
| 14 | LNF 100mg BID + 180ug PEG-IFN$\alpha$ QW | 3.34 | 35 | 2.12 | 56 | 2.73 | 2.70 |
| 15 | LNF 100mg BID + 180ug PEG-IFN$\alpha$ QW | 3.13 | 56 | 3.13 | 56 | 3.00 | 3.03 |
| Median [minimum, maximum] | | 3.13 [2.81,3.34] | 35 [14,56] | 2.64 [2.12,3.13] | 56 [56,56] | 2.73 [2.52,3.00] | 2.70 [2.56,3.03] |

**Table S4**. Listed are individual subject values for key pharmacokinetic parameters (Patients 1-15), where $C_{max}$ is the maximum observed plasma concentration, $T_{max}$ is the time of the maximum observed plasma concentration, $C_{last}$ is the last quantifiable plasma concentration, $T_{last}$ is the time of last quantifiable plasma concentration, $C_{avg}$ is the average plasma concentration over the dosing interval, and $C_{med}$ is the median plasma concentration over the dosing interval. Concentration is in log scale.

| Patient | $\kappa$ (log IU/mL) | $n$ | D($t_{end}$) (log IU/mL) | Fold increase in HBV DNA production rate |
|---|---|---|---|---|
| 1 | 3.42 | 1.1 | 6.87 | 1.0 |
| 2 | 5.2 | 1.1 | 5.85 | 1.2 |
| 3 | 6.56 | 7.7 | 6.6 | 1.5 |
| 4 | 4.2 | 11.1 | 3.87 | 5136.6 |
| 5 | 6.12 | 1.1 | 6.3 | 1.6 |
| 6 | 6.15 | 2.1 | 6.27 | 1.5 |
| 7 | 3.33 | 1.1 | 4.4 | 1.1 |
| 8 | 5 | 1.5 | 5.49 | 1.2 |
| 9 | 3.43 | 7.2 | 3.73 | 1.0 |
| 10 | 4.52 | 1.2 | 2.32 | 486.1 |
| 11 | 4.35 | 1.0 | 4.26 | 2.3 |
| 12 | 4.72 | 1.1 | 4.84 | 1.7 |
| 13 | 1.93 | 2.6 | 2.76 | 1.0 |
| 14 | 3.14 | 1.15 | 4 | 1.1 |
| 15 | 1.67 | 1.1 | 1.91 | 1.6 |
| Median [IQR] | 4.35 [3.33, 5.20] | 1.2 [1.1,2.4] | 4.40 [3.73, 6.27] | 1.5 [1.1,1.7] |

**Table S5:** Estimated fold increase in HBV DNA production rate following HDV decline at the patient level. The parameters $\kappa$ and *n* were borrowed from **Table 3** in the main manuscript whereas D($t_{end}$) denotes the HDV RNA level at the final time of our fittings, corresponding to the lowest observed serum HDV RNA level for each patient. Fold increase in HBV DNA production was estimated as $1 + (\kappa/D(t_{end} - \tau))^n$.